\documentclass{IEEEcsmag}

\usepackage[colorlinks,urlcolor=blue,linkcolor=blue,citecolor=blue]{hyperref}

\usepackage{upmath}

\usepackage{graphicx}
\usepackage[utf8]{inputenc}
\usepackage{indentfirst}
\usepackage{longtable}
\usepackage{hyperref}
\hypersetup{colorlinks=true,allcolors=blue}
\usepackage{hypcap}
\usepackage{svg}
\usepackage{amssymb}
\usepackage{pdflscape}
\usepackage[numbers]{natbib}
\usepackage{color, colortbl}
\usepackage{xcolor}
\definecolor{Gray}{gray}{0.9}

\usepackage{amsmath,amssymb}
\jvol{XX}
\jnum{XX}
\paper{8}
\jmonth{May/June}
\jname{IT Professional}
\pubyear{2019}

\graphicspath{ {./images/} }
\newcommand{\quotes}[1]{``#1''}

\begin{document}
	
	\title{Topic Modeling Based on Two-Step Flow Theory: Application to Tweets about Bitcoin}

	
	\author{\IEEEauthorblockN{Aos  Mulahuwaish\IEEEauthorrefmark{1},
			Matthew Loucks\IEEEauthorrefmark {1}, 
			Basheer Qolomany\IEEEauthorrefmark {2},  and Ala Al-Fuqaha\IEEEauthorrefmark {3}  
		}
		\IEEEauthorblockA{\IEEEauthorrefmark{1} Department of Computer Science and Information Systems, Saginaw Valley State University, University Center, USA}
		\IEEEauthorblockA{\IEEEauthorrefmark{2} Cyber Systems Department, University of Nebraska at Kearney, Kearney, USA}
			\IEEEauthorblockA{\IEEEauthorrefmark{3} Information and Computing Technologies (ICT) Division, College of Science and Engineering (CSE), Hamad Bin Khalifa University, Doha, Qatar}
		\thanks{Corresponding author: Aos Mulahuwaish (email: amulahuw@svsu.edu).}}

		\begin{abstract}
		 Digital cryptocurrencies such as Bitcoin have exploded in recent years in both popularity and value. By their novelty, cryptocurrencies tend to be both volatile and highly speculative. The capricious nature of these coins is helped facilitated by social media networks such as Twitter. However, not everyone's opinion matters equally, with most posts garnering little to no attention. Additionally, the majority of tweets are retweeted from popular posts. We must determine whose opinion matters and the difference between influential and non-influential users. This study separates these two groups and analyzes the differences between them. It uses Hypertext-induced Topic Selection (HITS) algorithm, which segregates the dataset based on influence. Topic modeling is then employed to uncover differences in each group's speech types and what group may best represent the entire community. We found differences in language and interest between these two groups regarding Bitcoin and that the opinion leaders of Twitter are not aligned with the majority of users. There were 2559 opinion leaders (0.72\% of users) who accounted for 80\% of the authority and the majority (99.28\%) users for the remaining 20\% out of a total of 355,139 users.  
		\end{abstract}

\maketitle



 \label{sec1}
\chapterinitial{The introduction} Due to the virtual nature of cryptocurrencies, much discussion occurs in online forums or social media platforms such as Twitter and Facebook. Places like this establish the general public opinion of cryptocurrencies and, consequently, to some extent, their price \cite {barber2012bitter, grinberg2011bitcoin, kim2016predicting, kim2016topiclens, marella2017bitcoin, reid2013analysis}. These networks, however, can comprise billions of users, with an unfathomable number of connections between them. To run any topic modeling or sentiment analysis algorithms across the entire network would be extraordinarily costly. This begs the question ― would it be possible to run topic modeling on a significantly smaller subset of the network and still yield similar results? It has been well established that authority, particularly on digital networks, tends to follow a Pareto distribution \cite{koch201180}. Additionally, most users tend to mimic opinion leaders, particularly on Twitter, where the built-in mechanism of retweeting facilitates the replication of one's opinions. By focusing on these hyper-concentrated authority centers on social media networks, it may be possible to use the opinion of the few to represent the entire community. 

When lower-end media users follow active media users who interpret the meaning of media messages and content, these are known as opinion leaders. These leaders are generally held in high esteem by those who accept their opinions. Opinion leadership originated from the theory founded by Elihu Katz and Paul Lazarsfeld, where it is theorized that there is a two-step flow of communication \cite{katz2017personal}. Significant contributors to the opinion leader concept include Berelson et al. \cite {riesman2020lonely}. The theory of opinion leadership is one of the multiple models that attempt to explain the diffusion of ideas, innovations, and commercial products.

Our research focused on analyzing the constructed network and finding the opinion leaders on a large and broader Twitter dataset, as it is more susceptible to a few's opinions than other platforms, analyzing the differences in language and interest between the opinion leaders and majority users regarding Bitcoin, and finding whether the opinion leaders of Twitter are not aligned with the majority of users, so that a small number of highly influential users (opinion leaders) can effectively represent an entire community's opinions. 

This paper does not consider the effects of mimicry and the extent to which it exists within these networks. For instance, after an opinion leader releases a tweet that corresponds to a specific topic, what is the precise effect on the community? From speculation, we would expect those users following them to begin to mimic their opinions — though we do not know for certain as our efforts were not focused on that question.

This paper and its motivations were heavily influenced by a previous study by Kang et al. \cite{kang2020whose} on a relative niche forum called bitcointalk.org. Given the recent hype surrounding cryptocurrency on mainstream social media platforms such as Twitter, it would be of great interest to replicate this previous study on a larger, broader dataset. We have seen how famous public figures such as Elon Musk have a tremendous capacity to influence and even embody public opinion. Additionally, given the difference in platform structure, Twitter may be more susceptible to a few's opinions than other platforms. For these reasons, we seek to replicate this original study using a Github fork and a few other modifications.

Also, in this paper, we used a different methodology for topic modeling techniques and the process in which a comparison is drawn between individual topics. In \cite{kang2020whose}, the Non-negative Matrix Factorization (NMF) has been used, while in this paper, we utilized the Latent Dirichlet Allocation (LDA)  due to its popularity and effectiveness in modeling the topics compared to the NMF. Additionally, in the \cite{kang2020whose}, the Hungarian Algorithm (also called the Munkres assignment algorithm or Kuhn-Munkres algorithm) was employed to individually match together the most closely related topics between opinion leaders and majority users. While in our paper, we reasoned, however, that not every topic extracted from the first group would have a 1-1 corresponding topic to the second group; thus, the matching process would be redundant or, at the very least, unproductive. Thus, we only considered qualitatively whether or not the topics from group to group were coherent. In addition, our methodology would be easily applicable to various social networks, such as finding the effectiveness of the influencers over social media for various aspects of the real world, such as political sentiment, divisiveness, fashion trends, music taste, or language changes.

\begin{figure*}[htbp]
	\center
	\includegraphics[width=7.6in]{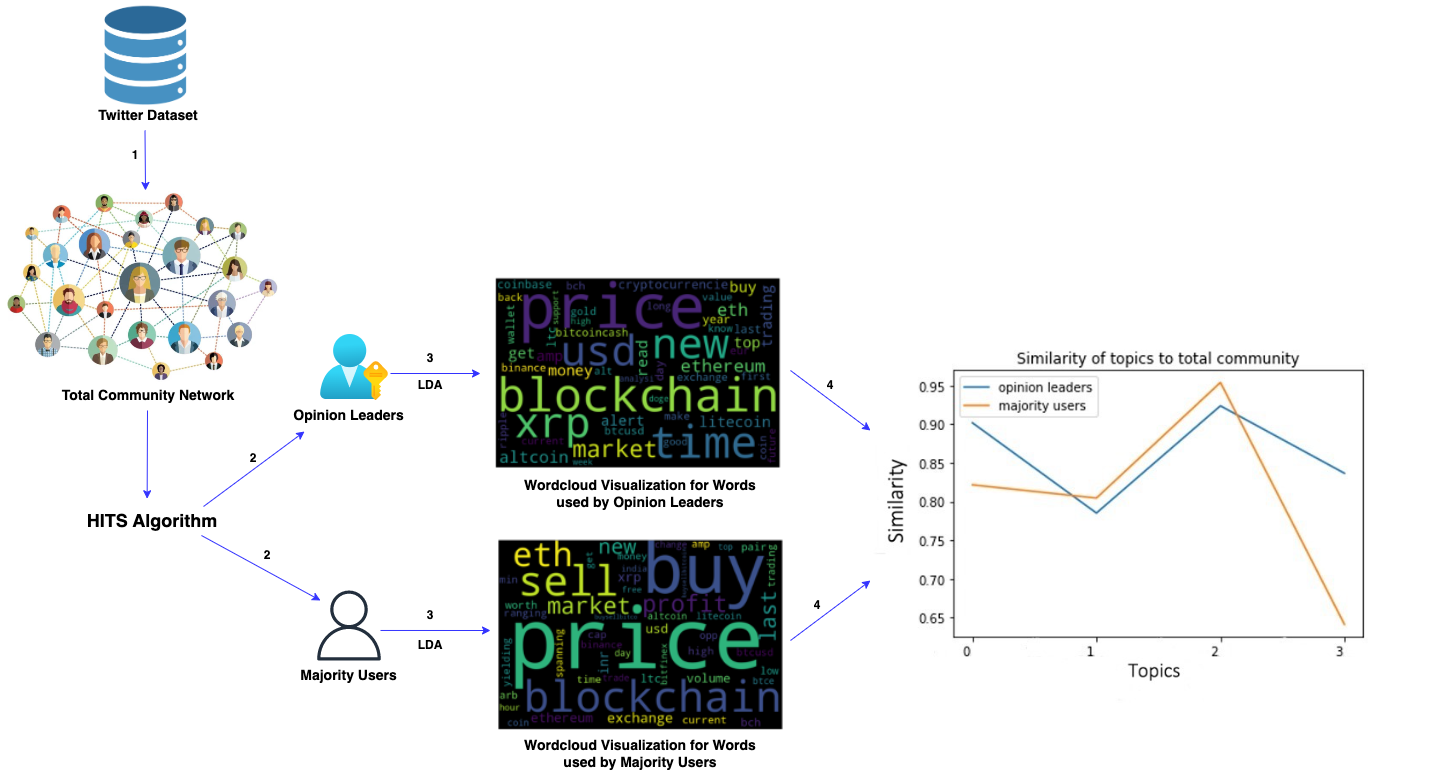}
	\caption{Steps of the proposed analysis approach.}
	\label{fig:fig1}
\end{figure*}

This study involves a few steps. In the first step, data is collected from Twitter using their API and preprocessed; the network is constructed using comments and retweets. In the second step, the network is segregated into two different groups using Hyper-text-induced Topic Selection (HITS) \cite{kleinberg1999authoritative}; these two groups are the opinion leaders and the majority users. In the third step, topic modeling is run on these two groups and the entire community. Finally, in the fourth step, we calculated the topic similarities between the two groups and the entire community to understand how well each constituent group represents the entire community. Figure \ref {fig:fig1} shows the steps conducted for our proposed analysis approach.

This paper is organized as follows. Section \ref {sec2} discusses related works. Section \ref {sec3} presents the data collection and preprocessing. Section \ref {sec4} presents network construction and analysis. Section \ref {sec5} presents the topic of modeling. Section \ref {sec6} presents the results of the work.Section \ref {sec7} presents the lessons learned. Finally, Section \ref {sec8} provides our conclusions.

\section{RELATED WORK} \label{sec2}
In this section, we will discuss some relevant, related works. 
	
Within prior studies on user opinions in online discussion forums and price fluctuation, sentimental and qualitative analyses were implemented. In addition, some of the said studies focused on particular phases where cryptocurrencies price rose. Multiple studies attempted to investigate how cryptocurrencies, such as Bitcoin price, were associated with users’ feelings and opinions by analyzing cryptocurrency forums. Some of these research works are the following:

Linton et al. \cite{linton2017dynamic} used the LDA to see how opinions are connected to large cryptocurrency events. The focus of this study was identifying major cryptocurrency events and catastrophes involving some sort of Ponzi scheme causing the market to crash. 30 was chosen for the number of topics due to their semantic coherence score. The dataset was obtained from scraping web data from bitcointalk.org, a cryptocurrency discussion forum. About 15 million posts were collected in total (2012 through 2016).   All data was scrapped from a single cryptocurrency forum which most likely exhibits a bias towards technologically savvy miners with a vested interest in the technology. No consideration was given to the influence of more prominent Bitcoin community members.

Abraham et al. \cite{abraham2018cryptocurrency} used sentiment analysis to predict price changes in cryptocurrencies (like Bitcoin and Ethereum). VADER algorithm is used to classify whether the text is positive or negative and its polarity. The datasets used in this work were collected from Google Trends and Twitter; as a result, there was no clear relationship between the general sentiments of tweets and price. Most tweets regarding cryptocurrency tend to be positive, and many papers use datasets in which the price of crypto increases, resulting in a biased dataset.

Rubio \cite{rubio2019analyzing} shifted focus away from exclusively Bitcoin and price to other cryptocurrencies and technical markers such as transaction speeds, smart contracts, and privacy. The researcher gathered tweets relating to over 25 top global cryptocurrencies from May 2018 to August 2018, checked if a categorization fits well for a group, detected prominent themes, and ran a prediction to see what would be used in the future, the LDA and MALLET algorithms were used in this work. As a result, the coherence scores plateaued at 18 topics, which gave a value of 18. The researcher then found the seven most common words per topic and visualized their frequencies in bar graphs. There were several themes between groups, including marketing, emotions, and group relations.

Nizzoli et al. \cite{nizzoli2020charting} present a chart of online cryptocurrencies across multiple platforms such as Twitter, Telegram, and Discord. They used a semi-supervised model to classify various social media platforms. They checked how prevalent pump-and-dump and Ponzi schemes are in each bot activity and mapped their extent using a popular Twitter bot detection framework. The authors used the Twitter Streaming API to gather tweets from 3822 hashtags. They then searched for additional Discord/Telegram links to crawl and scrape more data from those platforms. In total, they had 50M messages from March to May 2019. Data collected from Discord and Telegram was derived from Twitter, meaning any crypto communities on those sites that do not have any Twitter presence would not be included. As a result, Telegram showed to be replete with cryptocurrency scams; 56\% of crypto channels were fraudulent. Suspended ac-counts were also much more likely to be involved with crypto manipulation.

Atashian and Hrachya \cite{atashian2018sentiment} used the LDA to construct a predictive model for Bitcoin using sentiment analysis, as the internet primarily influences cryptocurrency. Data scraped from bitcointalk.org, a bitcoin forum, from April 23, 2011, to May 05, 2018. Used GDAX Python library to gather Bitcoin HOCV data from 2014 to 2018. Data was limited to niche esoteric cryptocurrency forums. This work Built a model with 15 topics total, with an R squared value of 0.28 in predicting bitcoin prices. The researchers also found adding LDA topic weights as a feature increases accuracy. Adding polarity and subjectivity scores from sentiment analysis also improved the R-value.

Bibi \cite{bibi2019cryptocurrency} used LDA to measure sentiment towards cryptocurrency worldwide. The Twitter dataset was used in this work; not much information was given regarding the dataset, algorithms, and methodology. As a result, Sweden and Denmark showed tremendously positive sentiment towards cryptocurrency, with Canada and South Korea being the most negative.

Kim et al. \cite {kim2017predicting} analyzed the current user opinions from online forums in Massive Multiplayer Online Games (MMOG) setting widely used around the world. This led to the proposition of a method for predicting the next day's fall and rise of the currency used in an MMOG environment. The prediction of the daily price fluctuations used in an MMOG environment was found by analyzing online forum users' opinions. The viability of predicting the fluctuation in the value of virtual currencies was shown by focusing on one of the most widely used MMOGs, the World of Warcraft game.

Kim et al. \cite {kim2017bitcoin} proposed a method founded on deep learning and established on user opinions on online forums to predict the fluctuation in the Bitcoin price and transactions. This method is viable for understanding a variety of cryptocurrencies in addition to Bitcoin and increasing the usability of these cryptocurrencies. 

Additionally, multiple studies sought to identify opinion leaders and/or find the main theme and maximize their network marketing effectiveness by analyzing community networks and users’ opinions. Some of these research works are the following:

Choi \cite {choi2015two} examined how a piece of information flowed in public forums that were social media based, whether opinion leaders appeared from this flow of information, and the characteristics that opinion leaders had in such forums. Two Twitter-based discussion groups focused on political discussions in South Korea were examined using network analysis and statistical measures, where it was concluded that opinion leaders were influential but not content creators. 

Ho et al. \cite {ho2016automatic} proposed an efficient approach to identify the opinion leader from a group discussion without analyzing syntactic and semantic features (which can add additional computing effort). The researchers proposed algorithms that evaluate the magnitude of participation and emotional expression during the speaking of each group member during the discussion. A well-trained model was tested on the single dataset, and a cross-dataset was obtained to recognize the opinion leader. This testing showed an accuracy of 94.68\% on the Berlin dataset, 76\% on the YouTube dataset, and 73.33\% on live discussion groups for identifying the opinion leader. 

Jiang et al. \cite {jiang2014detecting} designed a method for detecting an opinion leader based on an improved PageRank algorithm. This improved algorithm used link relevance to determine the degree of the link between users. Data was crawled from online communities and preprocessed, and then the weight matrix was constructed by calculating link relevance between users. The improved PageRank algorithm then ranked users and detected opinion leaders. Compared to the baselines, the results from this experiment showed that the proposed method effectively identified opinion leaders in online communities.

Wang et al. \cite {wang2016opinion} proposed a TopicSimilarRank algorithm for opinion leaders mining and interactive information based on the similarity of topics. This algorithm took into account text characteristics and user attributions in microblogs. This built links between users formed on user interaction information combined with topic similarity to construct a directed-weighted network. The idea of the vote in the PageRank algorithm to mine opinion leaders were also considered in this algorithm. Sina Weibo datasets were used for testing this algorithm, and the results showed that this algorithm had a better performance.

Zhao et al. \cite {zhao2018understanding} researched the influence power of opinion leaders and the interaction mechanism of a group of autonomous agents in an e-commerce community during forming group opinions. The social agents within a social network were divided into two subgroups opinion leaders and opinion followers. A new bounded confidence-based dynamic model for opinion leaders and followers was founded to simulate the opinion evolution of the group of agents. The results from this simulation concluded that enhancing opinion leaders’ credibility is key to maximizing the influence power within e-commerce.

In summary, from the literature review, a significant number of studies identified the opinion leaders on the network or analyzed users' sentiments on social media to predict the price fluctuation of the cryptocurrencies (like Bitcoin); these studies more likely used online materials to understand the fluctuation of Bitcoin prices and the volume of transactions to determine any relation. While in our research, we primarily focused on using network analysis and analyzing divergent behaviors and interests between opinion leaders and other majority users to find opinion leaders. In contrast to some of the aforementioned research works, our research did not emphasize predicting prices or volumes using causality analysis between price and word frequencies.

\section{DATA COLLECTION AND PREPROCESSING} \label{sec3}

\begin{figure}[h]
	\centering
	\includegraphics[width=2.1in]{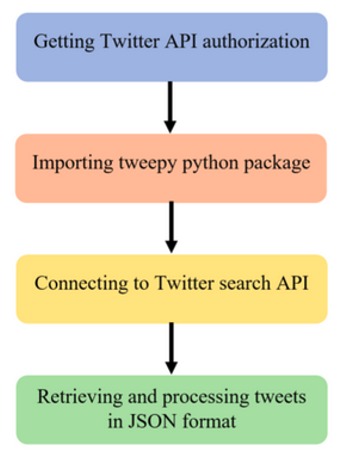}
	\caption{Process for collecting tweet data.}
	\label{fig:fig2}
\end{figure}

We collected 8 million cryptocurrency-related tweets for more than 100k users, scrapped from 2016-01-01 to 2019-03-29, leveraging the Twitter standard search application programming interface (API) and Tweepy Python library. A set of predefined search English keywords used such as Bitcoin or BTC, cryptocurrency. We extracted and stored the text and metadata of the tweets, such as timestamp, number of likes and retweets, hashtags, language, and user profile information, including user id, username, user location, number of followers, and number of friends. Figure \ref {fig:fig2} shows the data collection process. The first data collection phase is registering a Twitter application and obtaining a Twitter access key and token. The second phase is to import the Tweepy Python package and write the python script for accessing Twitter API. The third phase is connecting to Twitter search API and retrieving tweets using some cryptocurrency-related keywords. The last phase reads and processes the data to extract information on tags, agents, and locations for network construction and analysis. From this dataset, a subset of four million tweets was used. The specific break-down is as follows ― 2,753,808 posts, 533,924 comments, and 37,460 retweets.

Many modifications were made to the existing dataset before it was fed to the topic modeling algorithm. This process is crucial to ensure coherent results at the topic modeling stage. All non-English words were stripped as well as links and special characters. All text was lemmatized using WordNet and stripped of any stop words using the natural language toolkit (NLTK) library and a few additional custom stop words \cite{bird2006nltk}. All characters were converted to lowercase. Additionally, words with less than three characters and posts with less than five words were omitted.

\begin{figure*}[h]
	\center
	\includegraphics[width=5.0in]{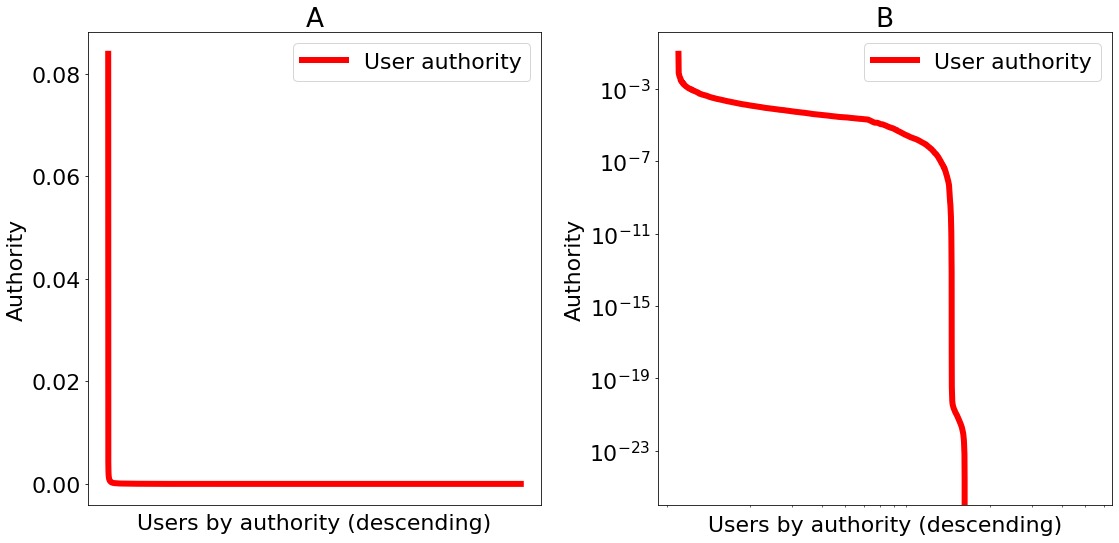}
	\caption{Graph of the distribution of authority (y-axis) sorted in descending order for both linear {\bf (A)} and log scale  {\bf (B)}. This figure includes both opinion leaders and majority users.}
	\label{fig:fig3}
\end{figure*} 

\section{NETWORK CONSTRUCTION AND ANALYSIS} \label {sec4}

For each of our networks, a node u is added to the graph for each user if and only if they have a post. For instance, if a user makes a comment but has no posts, they are not given a node. An edge V is drawn between two nodes {\textbf {\textit{ui}}} and {\textbf {\textit{uj}}} if the user corresponding with node {\textbf {\textit{ui}}} comments any post corresponding with node {\textbf {\textit{uj}}}. If a user comments on their post, no edge will be drawn ― in the context of graphs, no self-loops will be present. We did this because self-comments do not give significant insight into the relationship between other nodes and will thus only serve to add noise. Following this procedure, we constructed the entire community network.

Next, the HITS algorithm was used; the HITS algorithm works as a search engine to rank pages, and the algorithm uses hubs and authority to define relationships between pages. Hubs are highly valued lists for a given query. An authoritative page is one that many hubs link to, and a hub is a page that links to many authorities. We utilized the HITS algorithm to partition the total community network into two groups: majority users and opinion leaders. More specifically, this was done using the authority score from the HITS algorithm, which effectively shows the influence on neighboring nodes. The authority of a given node is defined as the sum of the hub scores from neighboring incoming nodes. The equations (1) and (2) for both authority and hubs are shown below:

\begin{equation}
Authority~({v_{u_{i}})} = \sum_{v_{u_{j}{\bf In} ({v_{u_{i}}) }}} hub~({v_{u_{j}})},
\end{equation}

\begin{equation}
	Hub~({v_{u_{i}})} = \sum_{v_{u_{j}{\bf Out} ({v_{u_{i}}) }}} authority~({v_{u_{j}})},
\end{equation}

We may define the authority of any given node $\bf(v_{u_i})$ as the sum of the hub scores from the set of adjacent nodes with {\bf incoming} connections. Similarly, the hub score of a node $\bf(v_{u_i})$ is defined as the sum of authority scores from the set of nodes with {\bf outgoing} connections. In practice, nodes with high authority are referenced frequently, and nodes with high hub scores link to many high authority nodes. For this work, we will focus on the authority score.

We define the opinion leader group as nodes whose authority sums to 80\% when considering the authorities of all users sorted in descending order. All other users are then to be considered the majority. As mentioned above, self-loops were omitted as they do not offer meaningful information about their relationship with other nodes. The content of retweets contained the text of the original tweet when fed through to the topic modeling algorithm. As far as implementation is concerned, we employed the Python NetworkX library to calculate the values for HITS. This is the converging algorithm and thus only ends when convergence is reached or a predefined limit is reached ― in our case, we set the maximum number of iterations to be 200.

The distribution of authority for the users in the network followed a Pareto distribution. The Pareto distribution, commonly referred to as the "80/20" rule," is commonly found in the allocation of capital and this case, influence. From this statistical observation, this work is based on ― most users hold little to no authority and thus have a negligible impact on the topics discussed on the network. This distribution can be seen as visualized in  Figure \ref {fig:fig3} . This figure points out the nature of the distribution of authority. We can see here that very few users possess the vast majority of the authority in this community. Also, we can see from the graph that most users have next to none. This distribution was plotted using linear and logarithmic scales for Figure \ref {fig:fig3} (A) and (B). This figure plots the distribution of authorities of users in descending order, where the x-axis is simply the index of each user in the list. The y-axis represents the authority. There were 2559 opinion leaders (0.72\% of users) who accounted for 80\% of the authority and the majority (99.28\%) users for the remaining 20\% out of a total of 355,139 users. These percentages were found by dividing the number of users in each subgroup by the total number of users.

\begin{figure*}[htbp]
\center
\includegraphics[width=3.5in]{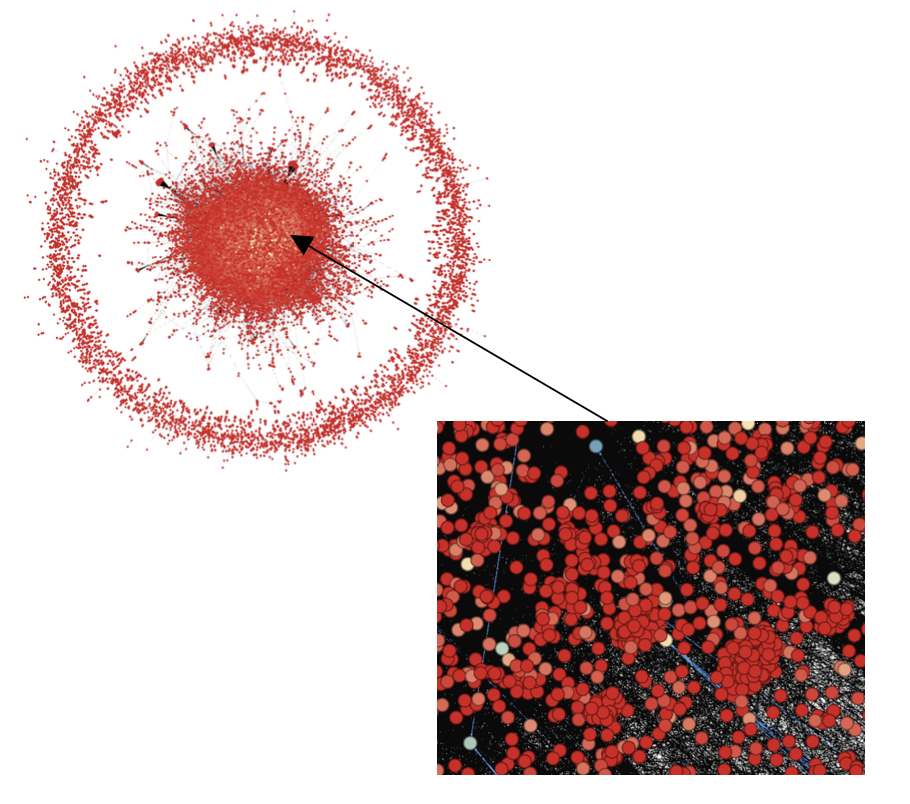}
\caption{Gephi visualization of complete network. This visualization helps illustrate the concentration of authority, demonstrated by the tiny cluster of blue central nodes surrounded by connecting majority users.}
\label{fig:fig4}
\end{figure*}

Additionally, we produced a network visualization using Gephi software \cite{bastian2009gephi}. Figure \ref {fig:fig4} visualize the entire community network. Red nodes represent the majority of users, and blue nodes represent opinion leaders. There is a gradient between the two based on authority scores. Black edges are comments, and blue edges are retweets. This was distributed using the ForceAtlas2 layout algorithm. 

\begin{figure}[h]
\centering \hspace*{-1.5cm}
\includegraphics[width=3.9in]{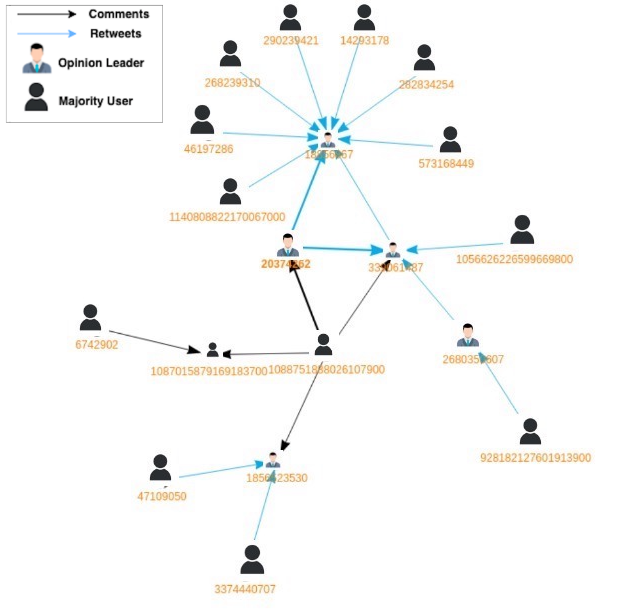}
\caption{A subset of a network illustrating the dynamics between opinion leaders. We can see from this example that opinion leaders are generally only hubs for other opinion leaders. Seldom is there an outgoing connection from an opinion leader to a majority user.}
\label{fig:fig5}
\end{figure}

Also, Figure \ref {fig:fig5} represents a subset of the entire network. Vector graphics, such as a person wearing a suit, show the nodes of the opinion leaders, while the other vector graphics show the nodes of the majority users; the numbers represent the account ID of each node. Black edges represent comments, and blue edges represent retweets. This example shows that opinion leaders garner substantial attention from other users, much more than most users.

\begin{table*}[]
\centering
\caption{Word frequencies for the opinion leaders and the majority users. The relative difference column shows by a factor how much larger the majority user percentage is than the opinion leaders.}
\begin{tabular}{|ccccc|}
	\hline
	\multicolumn{5}{|c|}{\bf Word Frequencies}                                 \\ \hline
	\multicolumn{1}{|c|}{}                  & \multicolumn{1}{c|}{}                                                                                 & \multicolumn{1}{c|}{\bf Opinion Leaders}                                                                    & \multicolumn{1}{c|}{\bf Majority Users}                                                                     &                                                                               \\ \hline
	\multicolumn{1}{|c|}{\bf Market}            & \multicolumn{1}{c|}{\begin{tabular}[c]{@{}c@{}}price\\ buy\\ sell\\ profit\\ invest\end{tabular}}     & \multicolumn{1}{c|}{\begin{tabular}[c]{@{}c@{}}0.85\%\\ 0.39\%\\ 0.19\%\\ 0.12\%\\ 0.04\%\end{tabular}} & \multicolumn{1}{c|}{\begin{tabular}[c]{@{}c@{}}1.89\%\\ 1.52\%\\ 1.29\%\\ 1.14\%\\ 0.06\%\end{tabular}} & \begin{tabular}[c]{@{}c@{}}1.22\\ 2.87\\ 5.74\\ 8.17\\ 0.53\end{tabular}      \\ \hline
	\multicolumn{1}{|c|}{\bf Technical}         & \multicolumn{1}{c|}{\begin{tabular}[c]{@{}c@{}}core\\ miner\\ network\\ node\\ protocol\end{tabular}} & \multicolumn{1}{c|}{\begin{tabular}[c]{@{}c@{}}0.07\%\\ 0.09\%\\ 0.16\%\\ 0.09\%\\ 0.05\%\end{tabular}} & \multicolumn{1}{c|}{\begin{tabular}[c]{@{}c@{}}0.02\%\\ 0.04\%\\ 0.09\%\\ 0.03\%\\ 0.02\%\end{tabular}} & \begin{tabular}[c]{@{}c@{}}-0.70\\ -0.53\\ -0.46\\ -0.67\\ -0.67\end{tabular} \\ \hline
\end{tabular}
\label {tab:table1}
\end{table*}

\section{TOPIC MODELING} \label{sec5}

Topic modeling is the process of deriving a statistical topic model from a collection of documents. It is a machine learning technique and is generally unsupervised. For example, topic modeling includes popular methods, including Latent Dirichlet Allocation (LDA) \cite{blei2003latent}, Non-negative Matrix Factorization (NMF) \cite{lee2000algorithms}, and Latent Semantic Analysis (LSA) \cite{landauer1998introduction}. In this paper, we decided to utilize LDA, a probabilistic algorithm, due to its popularity and effectiveness. The theory behind topic modeling is that each document is comprised of a mixture of various topics; the number of the topics is a mixture of different words. By categorizing documents by category, we can accurately and objectively group documents and gain the meaning behind them.

The LDA algorithm considers three hyperparameters: $\alpha$, $\beta$, and K. The $\alpha$ hyperparameter encodes the number of topics expected in each document. The $\beta$ hyperparameter lets it know the distribution of words for each topic in the document, and K defines the number of topics to use. Picking an accurate K value is crucial to an interpretable output from the algorithm. Choosing a high value may result in an output with topics with substantial overlap. Conversely, a low K would generalize too strongly and place most of the meaning behind a document into a smaller pool of topics.

\begin{figure*}[h]
	\center\hspace*{-0.9cm}
	\includegraphics[width=6.8in]{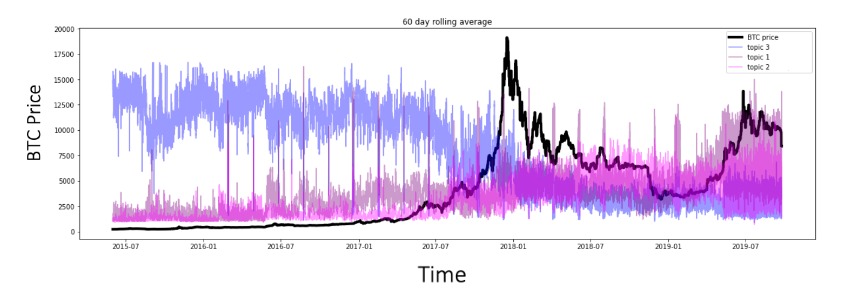}
	\caption{Bitcoin price and topics vs. time, the x-axis represents the time, and the y-axis represents the price of Bitcoin. The topic weights have been scaled to fit the graph instead of being on a 0-1 scale.}
	\label{fig:fig6}
\end{figure*}

\section{RESULTS AND DISCUSSION} \label{sec6}

The topic modeling algorithm method, Latent Dirichlet Allocation (LDA), derives the topics from our corpus of text. The implementation of this algorithm is handled by the popular data science package Scikit-learn \cite{pedregosa2011scikit}. This algorithm outputs a distribution of topics of what each document (tweet) is comprised of. We performed the LDA with a dataset of 3069704 tweets; the dataset timeframe is from 2015-06-01 to 2019-09-25 (YYYY-MM-DD). Table 1 shows the differences in the word frequencies between opinion leaders and majority users.

As we can see in Table \ref {tab:table1}, opinion leaders generally concern themselves more with cryptocurrency's technical aspects than the majority of users, who seem to be more interested in the price and profit. This result would make sense, as we would expect those with high levels of authority to possess greater expertise in the more technical aspects of cryptocurrencies. The words in Table \ref {tab:table1} are the most used in their respective categories. Also, we may define market as words pertaining to the financial aspects of Bitcoin, such as profit or price. Technical words, on the other hand, are regarded as words pertaining to the internal working or function of Bitcoin itself as a technology.

Figure \ref {fig:fig6} shows a graph of the Bitcoin price graphed alongside a scaled version of the topic weights smoothed out with a 60-day rolling average from the output of the total community LDA. Each document belongs to each topic to a certain extent — this is what the weight shows. Each color represents one of three top topics and runs from 2015 to 2019. The three topics are related to crypto, price, and trade, respectively. Each correlation contained a p-value and was calculated using the Pearson R score. As we might expect, different epochs of time yielded stronger correlations with a Bitcoin price. Between the period of December 2017 and April 2018, topic 3 yielded a correlation of -0.14. Between April 2018 and August 2018, the correlation of topic 3 dropped to -0.11. Finally, between January 2019 and May 2019, it went to -0.80. Topic 3 was chosen as an example due to its high correlation with the Bitcoin price. However, the other topics followed this pattern as well. We observed a general decline in correlation between topic weights and the price of Bitcoin over time. This may potentially be due to the rising shrewdness of investors ― as time goes on, investors may become less amenable to a suggestion by social media networks. Its mainstream adaption would lessen the influence of the subset of users with special technical knowledge.

The most important aspect of topic modeling is choosing the correct K value, or the number of topics to be selected. For this work, we opted not to use any heuristic-based approaches that utilize coherence scores and instead ran the model on varying topics, including 12, 8, and 4. We found the most interpretable results came from the output using four topics due to the minimized inter-topic similarity. Additionally, the human interpretability of each topic significantly improved upon lowering the number of topics.

After running LDA on the entire community, opinion leaders, and majority users, we used the cosine difference to calculate the degree of similarity between the groups. This allows us to numerically quantify the similarity between the groups by treating the corpus of text for each group as a word vector. A similarity score was calculated between the total community and opinion leaders and between the total community and majority users. Given these similarities, we could see which group was most similar to the total community and, thus, which one could represent it more appropriately.

\begin{figure*}[htbp]
	\center
	\includegraphics[width=5.0in]{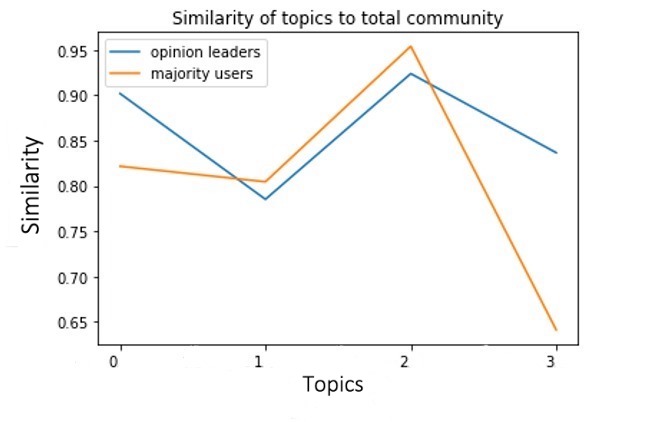}
	\caption{Graph of similarities to the total community for both opinion leaders (blue) and majority users (orange), the x-axis represents the topics, and the y-axis represents the cosine similarity to the topics of the entire community.}
	\label{fig:fig8}
\end{figure*}

Table \ref {tab:table2}  shows the output of the LDA algorithm for four topics for all three groups with the similarity scores. Additionally, Figure \ref {fig:fig8} shows the graph of the similarities for each topic. From the graph, it is evident that the opinion leader group's similarity closely matches and surpasses that of the majority user group. The similarity of the opinion leaders to the entire community is greater than that of the majority of users. Therefore, a small number of highly influential users (opinion leaders) can effectively represent an entire community’s opinions.

Table \ref {tab:table3} shows the values for the similarities between the groups. Similarity scores for each topic have been calculated for both opinion leaders and majority users. The average score has also been computed and is seen to be higher for the opinion leaders group, suggesting that, on average, the opinion leader group bears greater semblance to the entire community than the majority of users. Thus, this result also shows that the opinion leaders are more able to represent the entire community than the majority of users.

From the above discussions, we may notice that regarding the connection between BTC price and Tweets content, as with any matter concerning correlation followed by a claim, it is important first to ascertain that there is indeed a causal relationship that is not the product of mere happenstance — for it could be the case that the price of Bitcoin correlates strongly with any number of arbitrary datasets, so, how is this dataset of Bitcoin-related tweets from Twitter any different, we believe this topic has enough merit to warrant a separate study. It is evident that the influence of social media networks has tremendous potential to change markets. Look no further than the recent retail investor interest in GameStop or the unbelievable overvaluation of Tesla to get a hint of how much the value of some assets is driven by the current social media  \cite{fisch2022gamestop}. Given Bitcoin's speculative nature, its popularity amongst young adults (the primary consumers of social media), and the sheer volume of discussion that occurs about it online, we run with the assumption that there indeed exists some level of causality \cite{perrin2015social}.

In addition, if Twitter were to lose public trust, the methodology in the paper would still be easily applicable to different social networks. Our approach to this problem is not incompatible with the structure of other networks — we could have easily applied this to Facebook, Instagram, Reddit, etc., albeit with different results. Even if Twitter were to meet its demise, we would expect an equivalent to rise and fill its shoes. Our interest in Twitter, in particular, stems from its popularity amongst crypto influencers and its unique retweet mechanic. This makes it easier to differentiate between opinion leaders and majority users as there is a more stark dichotomy. If Twitter were to disappear, these opinion leaders would simply find a different platform on which we could easily reproduce the paper. Furthermore, while the paper's thesis (opinion leaders can effectively represent majority users) only extends so far as Twitter, replicating this study on other networks would confirm or disprove the thesis on a broader level which would shed greater insights on differences between these networks.

\begin{table*}
	\renewcommand{\arraystretch}{1.2} 
	\caption{LDA output for four topics for all three groups and similarity scores.} \label{tab:table2}
	\center
	\begin{tabular}{|c|p{35mm}|p{35mm}|p{13mm}|p{35mm}|p{13mm}|}
		\hline
	
	\bf {Topics} & \bf {Topics Keywords of Entire Community} & \bf {Topics Keywords of Opinion Leaders } & \bf {Opinion Leaders Similarity}  & \bf {Topics Keywords of Majority Users } & \bf {Majority Users Similarity }\\
		\hline
		\bf {Topic 0} &   bitcoin, price, usd, last, volume, exchange, market, hour, min, pair, profit, current, binance, arb, change, ranging, yielding, opps, spanning, eur & bitcoin, price, doge, current, day, year, crypto, dogecoin, bcash, month, amp, time, read, week, bch, market, since, hour, first, buy, get, last, back, good, volume, high, transaction, every, cryptocurrency, fee, eth, new,  low, hit, tweet, next, ago, ltc, bcashsv, block & 0.90195 & bitcoin, crypto, blockchain, cryptocurrency, ethereum, new, money, exchange, news, amp, get, mining, bank, world, currency, time, free, project, trading, cryptocurrencies, coin, future, ico, wallet, make, litecoin, gold, eth, token, digital, network, know, first, libra, good, payment, transaction, great, binance, facebook & 0.82172  \\
		\hline
		\bf {Topic 1} &  bitcoin, crypto, money, new, get, cryptocurrency, year, time, blockchain, amp, gold, make, know, day, world, good, future, bank, currency, news & bitcoin, xrp, ethereum, blockchain, crypto, time, litecoin, alert, cryptocurrencies, top, cryptocurrency, bitcoincash, ltc, trading, eth, price, ripple, binance, last, daysnews, vaultmex, market, cryptocurrencymarket, fintech, trx, altcoins, ico, buy, eos, altcoin, analysis, change, paypal, news, cryptonews, coinbase, newsoftheweek, monero, skrill, usd  & 0.78528 & bitcoin, eth, price, xrp, ltc, bch, volume, crypto, usd, cryptocurrency, ethereum, binance, hour, top, eos, litecoin, last, ripple, market, blockchain, change, trx, doge, bnb, etc, neo, sat, dash, ada, xlm, xmr, forecast, trading, bittrex, coin, average, link, news, eur, bitcoincash & 0.80462 \\
		\hline
	\bf {Topic 2}   & bitcoin, price, buy, sell, market, high, profit, worth, inr, current, low, crypto, btce, cap, time, bitfinex, btcusd, exchange, vircurex, analysis & bitcoin, usd, wallet, money, get, gold, eth, price, coinbase, hardware, value, eur, make, buy, amp, currency, altcoin, network, world, bank, know, time, apompliano, fiat, smart, work, year, mining, real, order, coin, good, secure, new, better, trezor, bsv, crypto, asset, digital  &  0.92394 & bitcoin, market, price, worth, high, low, cap, crypto, trade, last, btcusd, time, alt, usd, short, long, year, bull, next, day, bitmex, gmt, back, avg, bcoin, whale, hit, chart, bullish, get, look, week, bought, month, right, break, change, trading, good, move & 0.95417\\
		\hline
		\bf {Topic 3}  & bitcoin, crypto, blockchain, cryptocurrency, eth, ethereum, xrp, trading, ltc, bch, litecoin, free, binance, top, amp, ico, coin, ripple, join, cryptocurrencies  & bitcoin, crypto, price, market, cryptocurrency, read, news, btcusd, blockchain, bull, alt, new, bullish, support, trading, long, next, time, exchange, short, chart, high, amp, resistance, trader, back, future, usd, facebook, break, analyst, look, move, newsoftheweek, libra, analysis, bear, level, investor, craig   & 0.83673 &
	bitcoin, buy, sell, profit, price, inr, current, exchange, last, pair, min, ranging, arb, yielding, opps, spanning, btce, bitfinex, vircurex, btcusd, india, buysellbitcoin, buysellbitco, btceur, block, mtgox, edt, cest, utc, alert, hitbtc, est, cet, market, instantly, campbx, free, kraken, technical, trading & 0.64126 \\
		\hline
	\end{tabular}
\end{table*}

\begin{table}
	\renewcommand{\arraystretch}{1.2} 
	\caption{Average scores and topic similarities for opinion leaders and majority users.} \label{tab:table3}
	\centering
	\begin{tabular}{|c||c|c|c|c|c|}
		\hline
		\bf {Group} & 0 & 1 & 2 & 3 & \bf {Average}\\
		\hline
		\bf {Opinion Leader} & 0.90 & 0.77 & 0.92 &	0.85 & \bf {0.86}\\
		\hline
		\bf {Majority Users} & 0.83 & 0.81 & 0.96 &	0.65 & \bf {0.81}\\
		\hline
	\end{tabular}

\end{table}


\section {Lessons learned} \label{sec7}

We can conclude the following based on the results presented in this paper:
\begin{itemize}
	\item  According to the words shown in Table \ref {tab:table1}, opinion leaders use more technical words than the majority users, where the majority use words related to Bitcoin prices and profits. It implies that the opinion leaders would likely have a deeper understanding and experience of cryptocurrencies and Bitcoin.
	
	\item From Figure \ref {fig:fig6}, we can realize the correlation between the topic weights and the price of Bitcoin decreased over time. This may be because the investors become less interested in the suggestions from the social media networks. 
	
	\item We can see from Figure \ref {fig:fig8} that despite only accounting for a minute segment of the entire community, the opinion leaders possess an almost identical degree of similarity to the entire community as the majority users. We hypothesize that this is due to the influence of the opinion leaders on the rest of the community. This, coupled with Twitter's retweet feature, allows individuals' voices to propagate tremendously. 
	
	\item Based on the average scores and topic similarities shown in Table \ref {tab:table3}, we realized it is possible that a minuscule subset of a network may represent an entire community much more effectively than the majority users.
	
	\item The impact of any individual majority user is so low to almost considered non-existent compared to the effectiveness of the opinion leaders in the network.
	
	\item Measures of efficiency could improve certain analysis methods for user networks such as Twitter by only concentrating on the most influential voices.
	
	\item It is difficult to ascertain whether the price of Bitcoin drives tweets or tweets drive the price of Bitcoin. We know from isolated examples that tweets have the power to drive the price of a cryptocurrency, with the most salient examples being from figures such as Elon Musk, who routinely manipulates cryptocurrency markets and, according to Ante \cite {ante2023elon}, \quotes{non-negative tweets from Musk lead to significantly positive abnormal Bitcoin returns.} Given individual voices such as Musk can have such a significant individual influence on the price of Bitcoin; we may conclude that the composition of all such opinion leaders has a causal influence. Because of this, it is our assumption then that the confluence of these powerful voices does indeed, to some extent, influence the price of Bitcoin — though, to be clear, it is not certain to what extent this causality reaches.
	
	\item  Given what we have learned studying this topic, it would be interesting to see future studies applying the implied results of this paper. For instance, observing the opinion leaders of social media networks exclusively to observe their influence on various aspects of the real world, such as political sentiment, divisiveness, fashion trends, music taste, or language changes. It would be our assumption from the thesis of this paper that only the top voices would be relevant to the influence of these domains. It would also be interesting to observe situations where this theory falls apart — where the majority's voice outweighs that of opinion leaders. This dichotomy of opinion leader and majority user has the potential to provide an interesting lens through which to view the direction of future trends, not necessarily the price of a certain commodity. With a better understanding of how these two groups color the public narrative, we can better understand what lies ahead.

\end{itemize}

\section{CONCLUSION} \label{sec8}
This paper looked at the similarity between opinion leaders and majority users regarding Bitcoin on the social media platform Twitter. We conclude that a small number of highly influential users can effectively represent an entire community’s opinions. 

A more rigorous and deep-dive approach to the number of topics employed would be of great interest to improve upon this study. The results of this study were settled subjectively and did not incorporate any heuristic measures. Additionally, this study using specific languages or geographic locations may yield interesting results due to cultural differences.

Future research would benefit from a more rigorous form of topic-matching to ascertain a more concrete comparison. Also, discovering a quantifiable degree of mimicry between opinion leaders and majority users would be a great route to pursue. We may also expect opinion leaders to be influenced by their followers, following collectivized mimicry.

	\bibliography{ref} 


%

\section*{ACKNOWLEDGMENT}
	
This work was in part supported by Saginaw Valley State University.

\begin{IEEEbiography}
{Aos Mulahuwaish} received a Ph.D. in computer science from McMaster University, Hamilton, ON, Canada. He is currently an Assistant Professor in the computer science and information systems department at Saginaw Valley State University, MI, USA. His research interests include social media analytics, social cybersecurity, machine, and deep learning, metaheuristics, and fault tolerance system. He is a member of IEEE and ACM.
\end{IEEEbiography}

\begin{IEEEbiography}
{Matthew Loucks} is an undergraduate student currently studying computer science at Saginaw Valley State University. In his free time, he works as a part-time software developer and participates in his school's philosophy club. He enjoys reading, playing piano, and going to the gym. After graduating, Matthew is considering pursuing a master's either in computer science or artificial intelligence. 
\end{IEEEbiography}

\begin{IEEEbiography}
{Basheer Qolomany} [S'17-M'19] received the B.Sc. and M.Sc. degrees in Computer Science from the University of Mosul, Iraq, in 2008 and 2011, respectively, and the Ph.D. and second master’s en-route to Ph.D. degrees in Computer Science from Western Michigan University (WMU), Kalamazoo, MI, USA, in 2018. He has worked as a Lecturer with the Department of Computer Science, University of Duhok, Kurdistan region of Iraq, from 2011 to 2013; a Graduate Doctoral Assistant with the Department of Computer Science, WMU, from 2016 to 2018; and a visiting Assistant Professor at the Department of Computer Science, Kennesaw State University (KSU), Marietta, GA, USA, from 2018 to 2019. He is currently an Assistant Professor with the Department of Cyber Systems, University of Nebraska at Kearney (UNK), Kearney, NE, USA. His research interests include evolutionary computation, machine learning, deep learning, and big data analytics in support of population health and smart services. He is a member of IEEE and ACM. 
\end{IEEEbiography}

\begin{IEEEbiography}
	{Ala Al-Fuqaha} [S'00-M'04-SM'09] received Ph.D. degree in Computer Engineering and Networking from the University of Missouri-Kansas City, Kansas City, MO, USA, in 2004. He is currently a professor at Hamad Bin Khalifa University (HBKU). His research interests include the use of machine learning in general and deep learning in particular in support of the data-driven and self-driven management of large-scale deployments of IoT and smart city infrastructure and services, Wireless Vehicular Networks (VANETs), cooperation and spectrum access etiquette in cognitive radio networks, and management and planning of software defined networks (SDN). He is a senior member of the IEEE and an ABET Program Evaluator (PEV). He serves on editorial boards of multiple journals including IEEE Communications Letter and IEEE Network Magazine. He also served as chair, co-chair, and technical program committee member of multiple international conferences including IEEE VTC, IEEE Globecom, IEEE ICC, and IWCMC.
\end{IEEEbiography}

\end{document}